# Nowcasting Disaster Damage

Yury Kryvasheyeu [a,b], Haohui Chen [a,b], Nick Obradovich [c,d], Esteban Moro [e], Pascal Van Hentenryck [a,f], James Fowler [c,g], and Manuel Cebrian [a,h,*]

[a] National Information and Communications Technology Australia, Melbourne, Victoria 3003, Australia; [b] Faculty of Information Technology, Monash University, Melbourne, Victoria 3145, Australia; [c] Department of Political Science, University of California at San Diego, La Jolla, California 92093, USA; [d] Center for Marine Biodiversity and Conservation, Scripps Institution of Oceanography, La Jolla, California 92093, USA; [e] Department of Mathematics & GISC, Universidad Carlos III de Madrid, Madrid, Leganés 28911, Spain; [f] Research School of Computer Science, Australian National University, Canberra, ACT 0200, Australia; [g] Division of Medical Genetics, University of California at San Diego, La Jolla, CA 92093; [h] Department of Computer Science and Engineering, University of California at San Diego, La Jolla, California 92093, USA

[*] Corresponding author: MC, manuel.cebrian@nicta.com.au

Could social media data aid in disaster response and damage assessment? Countries face both an increasing frequency and intensity of natural disasters due to climate change. And during such events, citizens are turning to social media platforms for disaster-related communication and information. Social media improves situational awareness, facilitates dissemination of emergency information, enables early warning systems, and helps coordinate relief efforts. Additionally, spatiotemporal distribution of disaster-related messages helps with real-time monitoring and assessment of the disaster itself. Here we present a multiscale analysis of Twitter activity before, during, and after Hurricane Sandy. We examine the online response of 50 metropolitan areas of the United States and find a strong relationship between proximity to Sandy's path and hurricane-related social media activity. We show that real and perceived threats – together with the physical disaster effects – are directly observable through the intensity and composition of Twitter's message stream. We demonstrate that per-capita Twitter activity strongly correlates with the per-capita economic damage inflicted by the hurricane. Our findings suggest that massive online social networks can be used for rapid assessment ("nowcasting") of damage caused by a large-scale disaster.

social media | disaster management | damage assessment

## Introduction

Natural disasters are costly. They're costly in terms of property, in terms of political stability, and in terms of lives lost [1-3]. Unfortunately – due to climate change – natural disasters like hurricanes, floods, and tornadoes are also likely to become more common, more intense, and subsequently more costly in the future [4-7]. Developing rapid response tools to aid in adapting to these coming changes is critical [8].

As society faces this need, the use of social media on platforms like Facebook and Twitter is on the rise. Unlike traditional media, these platforms enable data collection on an unprecedented scale, documenting public reaction to events unfolding in both virtual and physical worlds. This makes social media platforms attractive large-scale laboratories for social science research [9-11]. Opportunities provided by social media are utilized in various domains including the economic [12], political, [13-16] and social [14, 17-21] sciences, as well as in public health [22, 23].

Because of their potential, the use of massive online social networks in disaster management has attracted significant public and research interest [24-26]. In particular, the micro-blogging platform Twitter has been especially useful during emergency events [27-29]. Twitter allows its users to share short 140-character messages and follow public messages from any other registered user. Such openness leads to a network topology characterized by a high number of accounts an average user follows, placing Twitter somewhere in-between a purely social and a purely informational network [30]. The information network properties of Twitter facilitate and accelerate the global spread of information; its social network properties ease access to geographically and personally relevant information, and the message length limit encourages informative exchange. These factors combine to make Twitter especially well suited for a fast-paced emergency environment.

Existing research on Twitter in emergency context studies its role in gathering and disseminating news [31, 32], the way it contributes to situational awareness [33, 34], practical aspects of classifying disaster messages, detecting events, and identifying messages from crisis regions [35-39]. Other research utilizes Twitter's network properties to devise sensor techniques for early awareness [40], to gauge the dynamics of societal response [41], and to crowdsource relief efforts [42].

More recently researchers have begun using social media platforms to derive information about disaster events themselves. For instance, the number of photographs uploaded to Flickr was shown to correlate strongly with physical variables that characterize natural disasters (atmospheric pressure during Hurricane Sandy) [43]. Although it is unclear what causes the link – external information, network effects, or direct observer effects – the correlation suggests that digital traces of a disaster can help measure its strength or impact. Based on a similar concept, other studies verify the link between the spatiotemporal distribution of tweets and the physical extent of floods [44] and the link between the prevalence of disaster-related tweets and a distribution of Hurricane Sandy damage predicted from modeling [45].

Here, we present a hierarchical multiscale analysis of disaster-related Twitter activity. We start at the national level and progressively employ finer spatial resolution of counties, and zip code tabulation areas. First, we examine how geographical and socio-cultural differences across the United States manifest through Twitter activity during a large-scale natural disaster – Hurricane Sandy. We investigate the response of cities to the hurricane and identify general features of disaster-related behavior on the community level. Second, we study the distribution of geo-located messages at the state level within the two most affected states (New Jersey and New York) and – for the first time – analyze the relationship between Twitter activity and the *ex-post* assessment of damage inflicted by the hurricane.

## Context of the study, data and methods

Hurricane Sandy was the largest hurricane of the 2012 season and one of the costliest disasters in the history of the United States. Sandy was a late season hurricane that formed on 22 October 2012 southwest of Jamaica, peaked in strength as a Category 3 hurricane over Cuba, passed the Bahamas, and continued to grow in size while moving northeast along the United States coast. The hurricane made its landfall on the continental United States at 12:00 UTC on 29 October 2012 near Brigantine, New Jersey with winds reaching 70 knots and the storm surge as high as 3.85 meters. According to the National Hurricane Center [46], Sandy caused 147 direct fatalities and is responsible for damage in excess of $50 billion, including 650 thousand destroyed or damaged buildings and over 8.5 million people left without power – some of them for weeks.

Both broadcast and online media extensively covered Hurricane Sandy, generating a large dataset of Twitter messages that became the basis for this study. The raw data comprises two distinct sets of messages. The first consists of messages with the hashtag "#sandy" posted between 15 October and 12 November 2012. This period precedes the formation of the hurricane and extends beyond its dissipation. The data includes the text of the messages and a range of additional information, such as message identifiers, user identifiers, followee counts, re-tweet statuses, self-reported or automatically detected location, timestamps, and sentiment scores. The second dataset has a similar structure and was collected within the same timeframe; however, instead of a hashtag, it includes all messages that contain one or more instances of specific keywords, considered to be relevant to the event and its consequences ("sandy", "hurricane", "storm", "superstorm", "flooding", "blackout", "gas", "power", "weather", "climate", etc. – see supplementary Table S1 for the full list). We obtained both datasets through the analytics company Topsy Labs [47], and reconstructed relationship graphs (list of followees for each user) using Twitter's API. Sentiment scores in raw data are measured by a proprietary algorithm from Topsy, which we additionally verify with free alternatives – Linguistic Inquiry and Word Count [48] and SentiStrength [49]. In total we have 52.55 million messages from 13.75 million unique users.

Since we are interested in a spatiotemporal analysis of Twitter activity, we focus exclusively on messages and users with known location. Only a fraction of the messages (about 1.2% for the hashtag dataset and 1.5% for the keywords dataset) are geo-tagged by Twitter. To expand the data, we include messages from users whose profile-listed address returns a match against the US Census Bureau Topologically Integrated Geographic Encoding and Referencing (TIGER) database. The resulting subset of geo-coded messages contains 9.7 million tweets from 2.2 million unique accounts.

We perform the analysis on the national and state levels. On the national level, we use cities as a natural – in terms of spatial extent and population size – basis for aggregation and comparison. Cities are important due to their dominant [50, 51] and increasing [52, 53] socio-economic role in all aspects of human life [54-56], both in the real world and online. Additionally, similarities or differences in the way cities react to a major natural disaster like Sandy are of interest to social scientists and climate adaptation policymakers alike [8, 57]. Our analysis covers the 50 most populous urban areas according to the 2010 US Census. On the state level we use a progressively finer granularity of counties, and finally zip code tabulation areas (ZCTA) to analyze the local distributions of Twitter activity and hurricane damage. On every level, we aggregate messages that have latitude and longitude falling within the boundaries of a respective region of interest (metropolitan area, county or ZCTA). The boundaries and population estimates of all administrative areas are determined by the 2010 US Census. Finally, for every city, we determine the shortest distance to the path of the hurricane [58], as an indication of the proximity to the disaster.

After aggregating the tweets by location, we use timestamps for temporal analysis. We allocate messages into non-overlapping bins of either 1- or 24-hour duration. Comparison metrics include the total number of active users, number of messages posted, classification of these messages into original and re-tweeted messages (including identification of the source as local or external to a particular community), and sentiment. Since the number of tweets originating from different urban or zip code areas varies greatly, we compare characteristics as normalized by the total count of distinct users for each area, active at any time between the start and end of the data collection period.

## Results

### Dynamics of Twitter activity across regions and hurricane-related topics

The messages studied here cover a range of keywords, with varying relevance to Hurricane Sandy. Because of this, we deal with three dimensions in our analysis – spatial, temporal and topical.

Figure 1 illustrates some of the characteristic features of Twitter activity. The pattern demonstrated by keywords strongly related to the hurricane ("sandy", "storm", "hurricane", "frankenstorm", etc.) is shown in Figure 1A – the number of messages slowly increases with a strong peak on the hurricane's landfall day, followed by a gradual decline in the tweet activity level. Geographically, the trend is similar almost everywhere, but the magnitude of the normalized response changes depending on the proximity to the hurricane.

An alternative way to summarize the activity is shown in the Figure 1B, where the normalized activity is presented as a two-dimensional heatmap. We rank cities by their proximity to the hurricane and words by their message count. At the peak of the disaster, event-related keywords rank higher and activity increases with proximity. Consequently, we see that the upper left corner of our city/topic matrix shows a high level of activity. In summary, as the disaster approaches and peaks in intensity, so does the normalized local Twitter response. Additionally, the content of the message stream changes and keywords most associated with the event dominate the agenda.

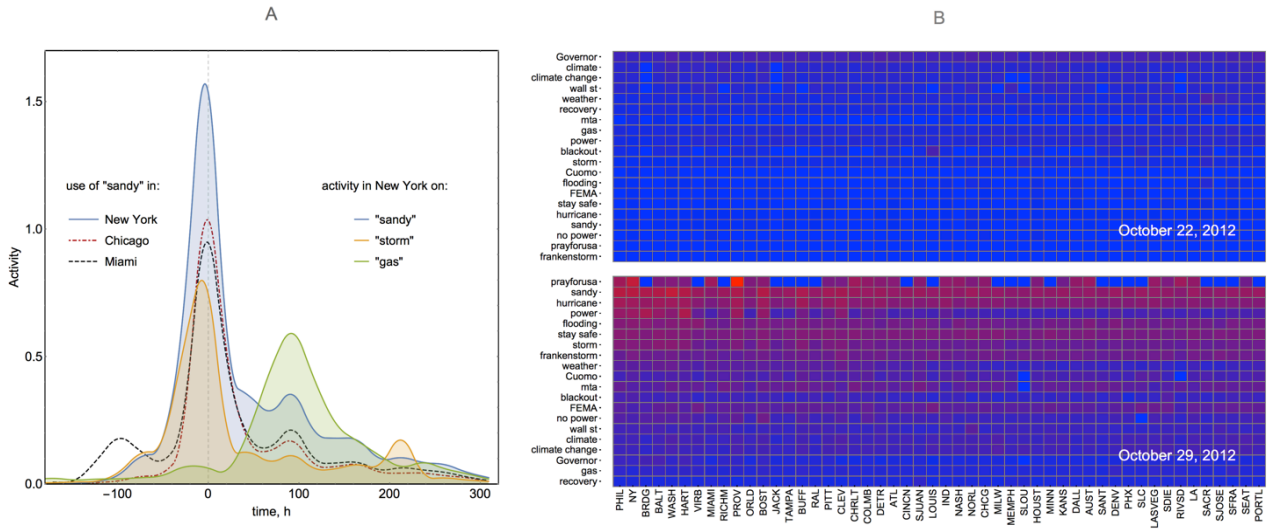

**Figure 1** Example of spatiotemporal evolution of Twitter activity across keywords. Panel A shows the geographical and topical variation of normalized activity (the number of daily messages divided by the number of local users active at any stage of the entire observation period). The horizontal axis is an offset in hours with respect to the landfall time of the hurricane (00:00 UTC on October 30, 2012). Activity on the hurricane-related words like "sandy" increases and reaches its peak in the day of landfall, and then gradually falls off. Qualitatively similar trends are observed everywhere, with distance to the path of the hurricane affecting the strength of the response (compare magnitudes of activity peaks between New York, Chicago and Miami). Different temporal patterns are exhibited by different keywords: "gas"-related discussion peaks with delay corresponding to the post-hurricane fuel shortages, and activity on "storm" has a secondary spike due to November "Nor'easter" storm. Panel B summarizes activity depending on both variables – topic and location. The color corresponds to the level of normalized activity (blue is low and red is high). In columns, places are ranked according to their proximity to the path of the hurricane (closest on the left, farthest on the right). In rows, words are ranked according to the total number of messages posted on the topic. Evolution of the event brings disaster-related words to the top of the agenda, with North East showing highest level of activity.

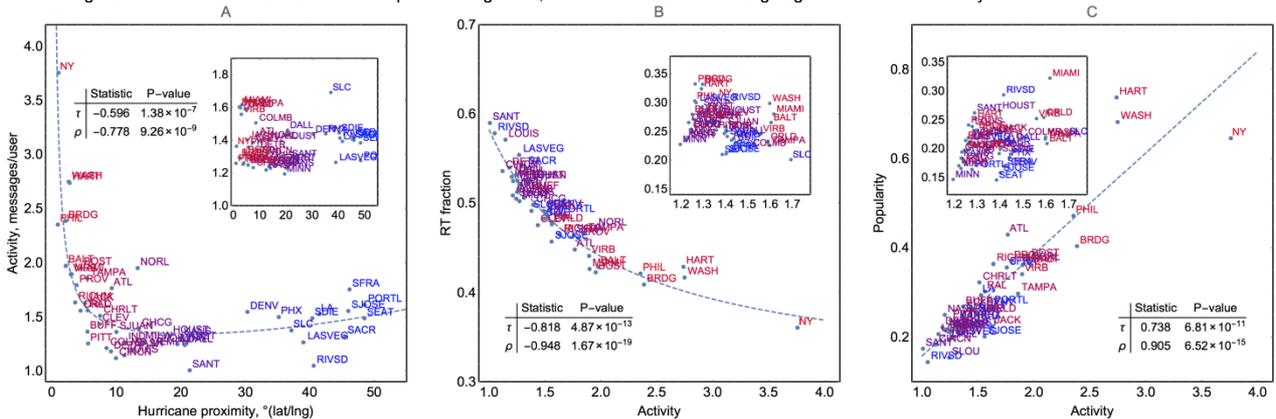

**Figure 2** Characteristic features of Twitter activity across locations (labeled by color according to hurricane proximity – blue is further and red closer to disaster). In all panels a primary plot shows results for messages with keyword "sandy", and an inset for keyword "weather", to contrast behavior between event-related and neutral words. A primary feature is the sharp decline of normalized activity as the distance between a location and the path of the hurricane increases (Panel A). After the distance exceeds 1200-1500 km its effect on the strength of response disappears. This trend may be caused by a combination of factors, with direct observation of disaster effects and perception of risk both increasing the tweet activity of the East Coast cities. Anxiety, anticipation, and risk perception evidently contribute to the magnitude of response, since many of the communities falling into the decreasing trend were not directly hit or were affected only marginally; while New Orleans for example shows significant tweeting level that reflects its historic experiences with damaging hurricanes, like Katrina. The re-tweet rate (Panel B) is inversely related to activity, with affected areas producing more original content. Popularity of the content created in the disaster area is also higher and therefore increases with activity as well (Panel C). None of the features discussed above are present for neutral words (see the insets in all panels).

When we aggregate our data over the period between 20 October and 12 November 2012, we find that tweet activity declines with increasing distance from the hurricane path up to 1500 km and is nearly constant for all places further away. These features are summarized in Figure 2A and supplementary Figure S1 (for all keywords). This relationship between proximity and activity level is a dominant feature, accompanied by two other relationships. The first one is an inverse relationship between content creation and consumption. The areas directly hit by – or close to – the disaster show a lower ratio of re-tweets (more original content) in the stream of messages generated, as can be seen in Figure 2B and Figure S2. The second relationship is between the activity and global popularity (count of messages that get re-tweeted, normalized by the local user count) of local content (Figure 2C and Figure S3), with content from affected areas attracting attention elsewhere. The activity-popularity (and to a lesser degree activity-originality) relationship is very strong for the event-related keywords, but virtually absent for neutral or more general keywords. We illustrate this in Figure 2 by the inset plots for keyword "weather" – a general word that is used frequently and is not necessarily associated with extreme weather events, even when such events take place.

Direct relationship between online activity and proximity to the hurricane naturally raises the question of factors that stimulate such an activity. Is it extensive media coverage, or perception of risk, or witnessing hurricane's meteorological effects (winds, precipitation, storm surge) and damage (power and fuel shortages, flooding, loss of personal property, casualties)? The latter, especially the extent to which quantifiable properties of online activity – recorded during and shortly after disaster – reflect the severity of disaster-related damage, is especially interesting from the disaster management point of view. Real-time analysis of online activity as a predictor of damage would be a valuable tool in optimizing allocation of limited emergency and recovery resources, and may complement other predictive models used in the joint assessment and recovery of damaged infrastructures [59]. Therefore, we investigate if the damage to property across the most severely hurricane-affected regions correlates with the recorded Twitter activity.

**Damage**

Since the hurricane damage was mostly confined to several states, we perform damage analysis at finer spatial granularity by looking at counties and ZCTAs. We examine both aggregation levels to determine the limits of spatial resolution achievable in such a "nowcasting" technique.

Two primary sources of data contribute to our estimate of damage. The first is data on FEMA Household Assistance grants to homeowners and renters [60]. These grants are provided to relieve the hardship of households exposed to disaster and to enable bringing the original property back to a habitable condition. The second source is data on insurance claims associated with Hurricane Sandy [61, 62], including National Flood Insurance, residential, commercial, vehicle and marine insurance claims. We use these indicators as both are expressed in monetary terms and are reported by individuals, rather than administrative entities like municipalities. A more holistic index of community hardship could be developed, like the one in [63] taking into account other metrics: the number of people served at shelters, effects of power loss (using the number of days schools were closed as a proxy), gas shortages (through the number of calls to State Emergency Hotline from gas stations) and FEMA Public Assistance grants to help with municipal infrastructure. Although such methodology gives a broader picture of the hardship on the ground, the metrics involved do not have a standard way of measurement and do not share a common unit to be integrated together, which leaves certain freedom to assign arbitrary weights to each contributing factor. To avoid this ambiguity, we only include the data reported by individuals and measured directly as monetary loss.

We analyze the damage estimates, aggregated within either counties or zip code tabulation areas, against Twitter activity in the same boundaries. The data available on damage allows us to look at several aspects, including the total damage claimed, the total damage covered by FEMA and insurance, the number of applications and successful applications, and severity categories based on the cost. We look at the relationship between normalized quantities – per-capita Twitter message count and per-capita damage – to avoid correlations artificially induced by population counts (more populous areas produce higher message counts and experience higher damage). To estimate whether activity quantitatively reflects the severity of the disaster, we test the independence of two distributions: activity versus damage. We consider activity on the core set of messages strongly associated with the hurricane (see supplementary Table S2 for the rankings and Table S3 for the results across all keywords).

The estimate of damage is a snapshot from November 2014, while activity varies significantly over the data collection period. Both in the interest of capturing predictive capacity and in a practical attempt to determine the best analysis window to get the strongest predictive effect, we calculate the correlations on a daily basis between the 22nd of October and the 12th of November. In addition to examining the activity-to-damage correlation, we also check the sentiment-to-damage correlation. Previous studies [40] suggest that a drop of the average sentiment in an area may indicate an emergency, and we aim to verify whether the sentiment also serves as a quantitative predictor of damage.

The correlation coefficient dynamics is presented in Figure 3. Because we discard inactive areas – ZCTAs with no messages posted during an analysis period – the length of vectors subject to an independence test varies over time, and we choose to ignore correlation coefficients earlier than October 22 and later than November 11. Within this period, we consistently have more than 200 active ZCTAs, see Figure 3A. Figure 3B shows that the rank correlation coefficients are moderately positive, indicating a weak correlation.

The correlation is present for several days before the landfall, which might reflect *a-priori* knowledge of local hurricane vulnerability based on historical experience within particular areas and obvious risk factors like proximity to the shoreline. This positive correlation decreases on the day of landfall across all correlation measures. Despite the highest total count of messages, the peak of the disaster has the weakest damage-predictive power. However, in the following two days, the activity-to-damage correlation steadily increases. From the third day onwards, it fluctuates around a moderate level (0.25 Kendall $\tau$ and 0.35 Spearman's $\rho$). We examined these trends both combining all data as well as for different keywords separately, without much of a difference in pattern or magnitude of coefficients.

Arguably, this trend – a drop on the day of the hurricane followed by a steady increase in the relationship between activity and damage – could be explained by the universally high tweet activity on the day of hurricane landfall, fuelled not only by the severity of the storm, but also by widespread coverage of the hurricane in all forms of the media. In places that were spared significant consequences of the hurricane, the interest of the public quickly diminishes. But in affected areas, the topic persistently remains at the top of the agenda and makes post-event activity an indicator of the damage caused by the hurricane.

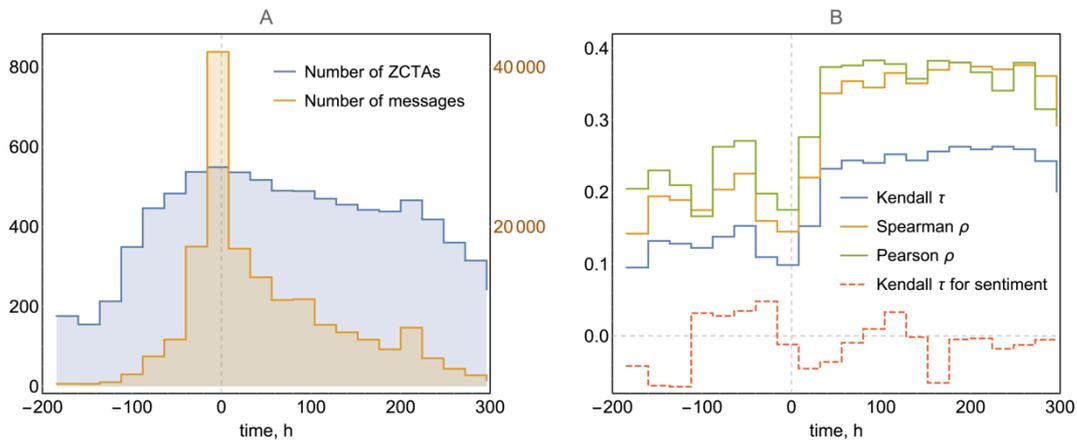

**Figure 3** The predictive capacity of Hurricane Sandy's digital traces. The horizontal axis is an offset in hours with respect to the landfall time of the hurricane (00:00 UTC on October 30, 2012). Panel A shows the number of messages as a function of time (labeled on the secondary y-axis on the right) and the number of "active" – with at least one message posted – zipcode tabulation areas (labeled on the primary y-axis on the left). Panel B shows evolution of the rank correlation coefficients between the normalized per-capita activity (number of the original messages divided by the population of a corresponding ZTCA) and per-capita damage (comprised of FEMA Individual Assistance grants and Sandy-related insurance claims). Additionally, the dashed trend shows Kendall rank correlations between average sentiment and per-capita damage. The correlation rises from pre-landfall to post-landfall stage of the hurricane, with a drop on the landfall day. We conclude that the post-disaster stage, or persistent activity on the topic in the immediate aftermath of an event, is a good predictor of damage inflicted locally. Strength of the average sentiment of tweets does not seem to be a good predictor, at least at this level of spatial granularity (ZCTA resolution).

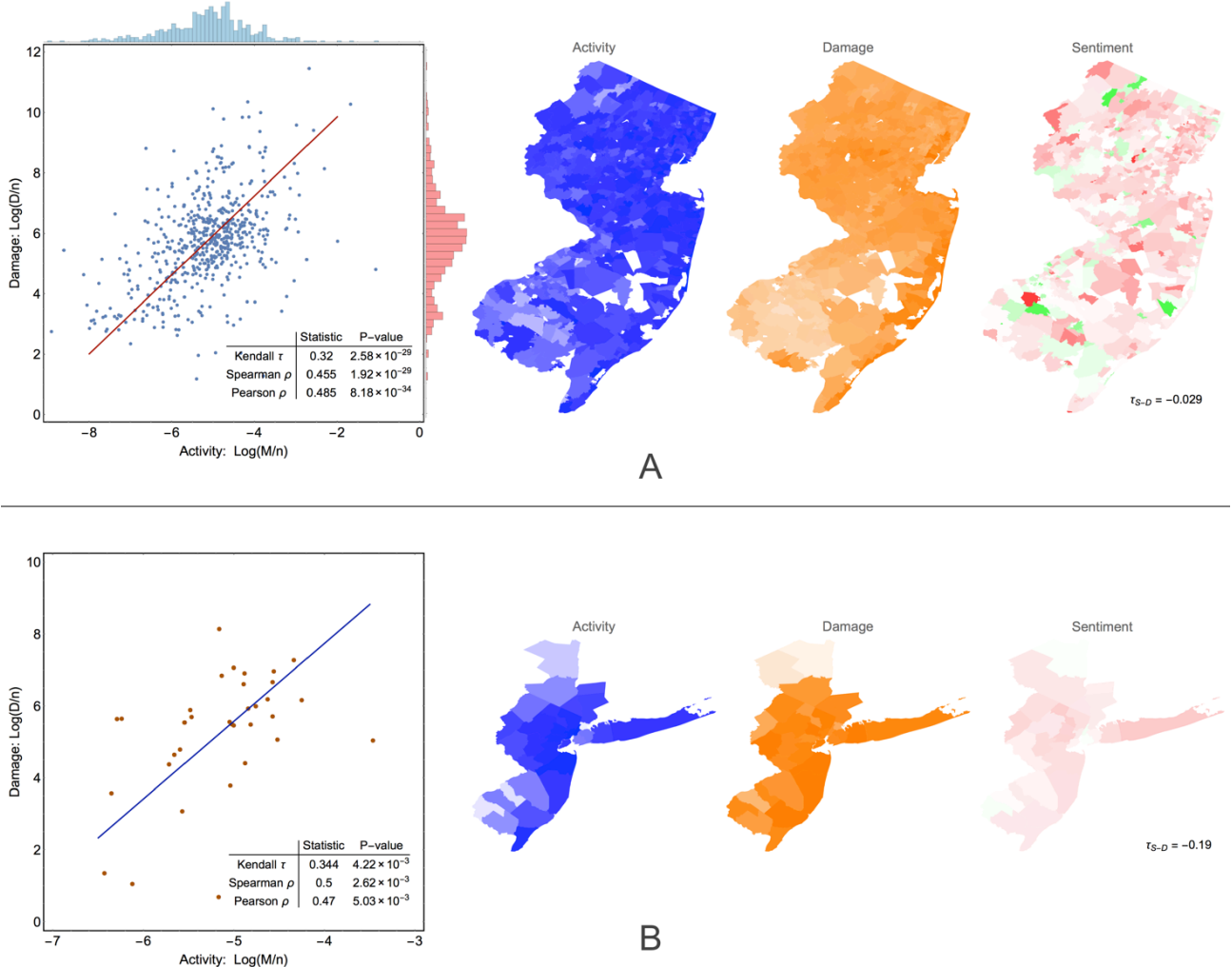

**Figure 4** Spatial distributions and mutual correlations between Hurricane Sandy damage, Twitter activity, and average sentiment of tweets. Correlations between per capita Twitter activity and damage are illustrated on ZCTA-level (Panel A) and county level (Panel B). In this normalized form both variables follow a quasi log-normal distribution (see the histograms along the axes of scatter plot in Panel A). There is a moderately strong positive correlation between post-landfall activity and damage, with rank coefficients approximately equal at either level of resolution (see inset tables in scatter plots in panels A and B for exact statistics and P-values). ZCTA analysis covers only the state of New Jersey and spatial distribution shows that both activity and damage reach highest levels along the coast and in a densely populated metropolitan area around New York City. There is no correlation between sentiment and damage at the level of ZCTA granularity (Kendall-tau rank coefficient is -2.52 10-3). County-level analysis in Panel B shows that the strength of the activity-to-damage correlation holds, and there is also weakly negative correlation between sentiment and damage.

Focusing on the period where the relationship between activity and damage is the strongest – between October 31 and November 12 – we measure rank correlation coefficients for all ZCTAs in New Jersey and for selected counties in New Jersey and New York. Results are summarized in Figure 4 and Figure S4. ZCTA-based distributions of per-capita activity and per-capita damage are approximately log-normal, with histograms shown in Figure 4A. The Kendall rank correlation reaches 0.32, the Spearman rank correlation 0.46, and the Pearson correlation coefficient approaches 0.5. Analysis on the basis of counties (Figure 4B) reveals similar results: Kendall $\tau = 0.34$, Spearman $\rho = 0.5$ and Pearson $\rho = 0.47$ for 34 counties across New Jersey and New York. All measures are statistically significant with *P*-values below 0.05 and indicate a moderate positive correlation between damage and tweet activity. Spatial distributions confirm the relationship, with pronounced concentration of both damage and normalized activity along the coastline of New Jersey. Alternative normalization (by Twitter user count instead of actual population) does not alter the strength of the correlation – see supplementary Table S4.

Following Guan and Chen [45], we also analyze the relationship between Twitter activity and damage estimates produced by the FEMA Modeling Task Force (based on the Hazus-MH model of hurricane wind and storm surge damage to housing and infrastructure). This approach results in correlations of similar strength (Kendall $\tau = 0.34$, Spearman $\rho = 0.51$ and Pearson $\rho = 0.41$). Comparison of the alternative damage estimates and their negligible effect on the strength of the observed activity-damage correlation is summarized in the supplementary tables S5 and S6.

Our previous study [40] suggested that the negative average sentiment may indicate an emergency situation, based on the fact that the sentiment experiences a drop for a sustained period of time before and after the landfall of Hurricane Sandy. Here, we re-examine the sentiment-damage relationship and find that daily ranking correlation coefficients oscillate around zero for the entire observation period (see Figure 3B). Within the most favorable prediction window (October 31 to November 12) Kendall $\tau = -0.029$ ($P = 0.31$), suggesting independence of the underlying distributions, or that analysis at ZCTA-resolution is under-powered. Change of spatial resolution from ZCTAs to counties results in a more definitive relationship with $\tau = -0.19$ ($P = 0.11$), and normalization by Twitter user count yields more significant results yet – $\tau = -0.23$ ($P = 0.06$) – confirming our previous findings and making sentiment weakly predictive of damage (see supplementary Table S7 for the summary of results).

## Discussion and conclusions

We found that Twitter activity during a large-scale natural disaster – in this instance Hurricane Sandy – is related to the proximity of the region to the path of the hurricane. Activity drops as the distance from the hurricane increases and, after a distance of approximately 1200 – 1500 km, the influence of proximity disappears. Additionally, the areas close to the disaster have a lower fraction of re-tweets (more original content) and generate more interest by producing content popular elsewhere – the findings confirmed by other studies [37].

In the first study of its kind based on the actual *ex-post* damage assessments, we demonstrated that the per-capita number of Twitter messages corresponds directly to disaster-inflicted monetary damage. The correlation is especially pronounced for persistent post-disaster activity and is weakest at the peak of the disaster. We established that per-capita activity and per-capita damage both have approximately log-normal distribution and Pearson correlation coefficient between the two can reach 0.5 for a carefully selected observation period in the aftermath of the landfall. This makes social media a viable platform for preliminary rapid damage assessment in the chaotic time immediately after disaster. Our results suggest that, during a disaster, officials should pay attention to the normalized activity levels, rates of original content creation, and rates of content rebroadcast to identify in real time the hardest hit areas. Immediately after a disaster they should focus on persistence in activity levels to assess which areas are likely to need the most assistance.

The role of proximity as the primary factor that explains activity suggests that individuals realistically assess danger based on personal experiences [64], and their level of interest is moderated accordingly. The cutoff in the activity-to-distance relationship is on the same order of magnitude as the footprint of a large atmospheric system, indicating that once people feel safe where they are, the level of engagement is uniform and most likely depends on the intensity of media coverage. Activity within the zone of the disaster sharply rises with proximity to its epicenter, probably due to a combination of factors including heightened anxiety, sense of direct relevance, and observation of the associated effects (wind and precipitation, physical damage). Our findings echo other studies, like the correlation of the number of Flickr photos tagged "#sandy" with atmospheric pressure over New Jersey, emphasizing that online activity is increasing with the intensity of the event [43]. What is striking however, with all the different factors that motivate people to tweet, is that a simple normalized measure of this activity – per-capita number of messages – serves as an efficient assessment tool for the physical damage caused by the disaster.

Damage forecasts issued by FEMA's Modeling Task Force rely on the sophisticated multi-hazard modeling. Although these forecasts are timely (generated before or immediately after a disaster), their verification with aerial imagery and physical site inspections is resource and time-consuming. Social media nowcasting provides additional low-cost tool in the arsenal of authorities to expedite the allocation of relief funds. In the long term, the technique can be used to check the integrity of damage assessment process itself, especially in light of protracted settlement timeframes and allegations of irregularities that recently prompted a blanket review of all insurance claims by FEMA [65]. It can be used to inform stochastic optimization algorithms for the joint assessment and repair of complex infrastructures such as power systems [59].

The correlation we observed is not definitive in its strength, and care should be taken in the attempt to devise practical applications, however we believe there is potential to fine-tune the method. More robust estimates of damage through other data sources – for instance an in-clusion of municipal losses and non-monetary indicators like power losses and emergency shelters statistics [63] - may reinforce the relationship. Composite metrics that combine per-capita activity with other properties of Twitter message stream, e.g. fraction of disaster related tweets [45] and sentiment (provided that activity is high and the volume of data is sufficient for sentiment to be predictive), may prove even more sensitive to damage. More broadly, our study suggests that the distribution of per-capita online activity on a specific topic has potential to describe and quantify other natural, economic or cultural phenomena.

**Acknowledgements** Authors would like to thank Colleen O'Dea (NJ Spotlight, www.njspotlight.com) for helpful suggestions and advice on obtaining the data for Hurricane Sandy damage, Federal Emergency Management Agency, New Jersey State Department of Banking and Insurance, and New York State Department of Financial Services. Yury Kryvasheyeu, Haohui Chen, Pascal Van Hentenryck and Manuel Cebrian acknowledge the support of the Australian Government represented by the Department of Broadband, Communications and Digital Economy and the Australian Research Council through the ICT Centre of Excellence program. This research was also supported by the Army Research Laboratory under Cooperative Agreement Numbers W911NF-09-2-0053 and W911NF-11-1-0363, the National Science Foundation under grant 0905645, and DARPA/Lockheed Martin Guard Dog Programme under PO 4100149822.

# Nowcasting Disaster Damage: Supplementary Information.

Yury Kryvasheyeu [a,b], Haohui Chen [a,b], Nick Obradovich [c,d], Esteban Moro [e], Pascal Van Hentenryck [a,f], James Fowler [c,g], and Manuel Cebrian [a,h,*]

[a] National Information and Communications Technology Australia, Melbourne, Victoria 3003, Australia; [b] Faculty of Information Technology, Monash University, Melbourne, Victoria 3145, Australia; [c] Department of Political Science, University of California at San Diego, La Jolla, California 92093, USA; [d] Center for Marine Biodiversity and Conservation, Scripps Institution of Oceanography, La Jolla, California 92093, USA; [e] Department of Mathematics & GISC, Universidad Carlos III de Madrid, Madrid, Leganés 28911, Spain; [f] Research School of Computer Science, Australian National University, Canberra, ACT 0200, Australia; [g] Division of Medical Genetics, University of California at San Diego, La Jolla, CA 92093; [h] Department of Computer Science and Engineering, University of California at San Diego, La Jolla, California 92093, USA

[*] Corresponding author: MC, manuel.cebrian@nicta.com.au

Could social media data aid in disaster response and damage assessment? Countries face both an increasing frequency and intensity of natural disasters due to climate change. And during such events, citizens are turning to social media platforms for disaster-related communication and information. Social media improves situational awareness, facilitates dissemination of emergency information, enables early warning systems, and helps coordinate relief efforts. Additionally, spatiotemporal distribution of disaster-related messages helps with real-time monitoring and assessment of the disaster itself. Here we present a multiscale analysis of Twitter activity before, during, and after Hurricane Sandy. We examine the online response of 50 metropolitan areas of the United States and find a strong relationship between proximity to Sandy's path and hurricane-related social media activity. We show that real and perceived threats – together with the physical disaster effects – are directly observable through the intensity and composition of Twitter's message stream. We demonstrate that per-capita Twitter activity strongly correlates with the per-capita economic damage inflicted by the hurricane. Our findings suggest that massive online social networks can be used for rapid assessment ("nowcasting") of damage caused by a large-scale disaster.

**Table S1** List of keywords included in the analysis, with their corresponding message counts.

| Keyword | Count |
|---|---|
| power | 4 825 717 |
| sandy | 4 745 099 |
| hurricane | 4 680 290 |
| weather | 3 333 025 |
| storm | 2 555 196 |
| gas | 1 991 524 |
| Governor | 498 135 |
| stay safe | 484 732 |
| recovery | 431 591 |
| climate | 420 217 |
| FEMA | 329 789 |
| flooding | 264 132 |
| no power | 261 998 |
| climate change | 236 009 |
| wall st | 233 411 |
| blackout | 213 520 |
| mta | 206 504 |
| frankenstorm | 205 467 |
| Cuomo | 92 014 |
| prayforusa | 91 293 |

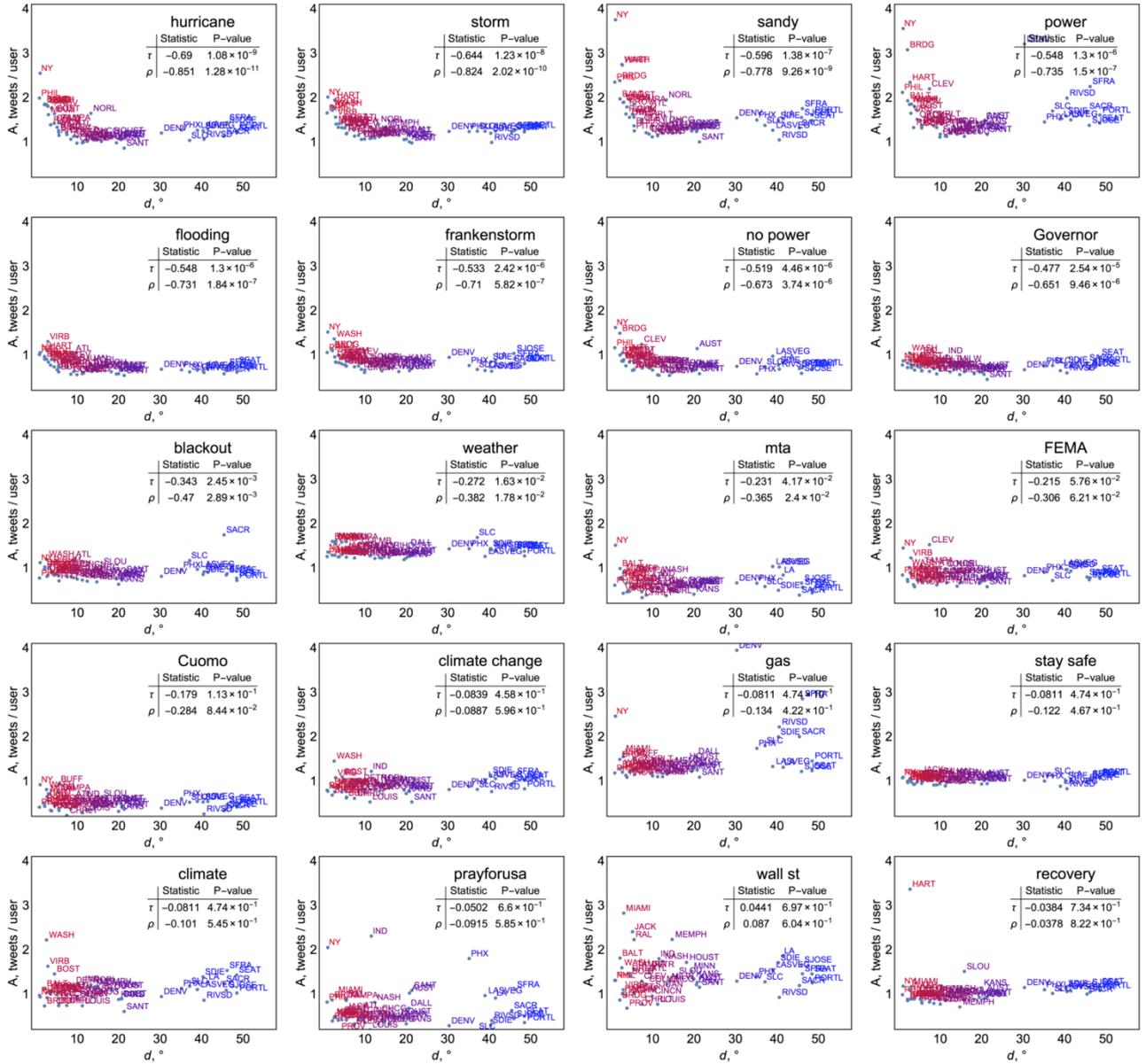

**Figure S1** Normalized local activity on the topic as a function of distance to the hurricane path. For the words strongly related to Hurricane Sandy (top row) activity decreases with distance, and after the distance of 10-15 degrees (1200 – 1500 km) proximity does not affect the level of activity. Because if this, and also because for some words ("gas", "power", etc.) trends for East and West coasts differ, we use rank correlation for East coast cities as a measure of relevance. The values of these correlation coefficients between activity and distance are shown in insets.

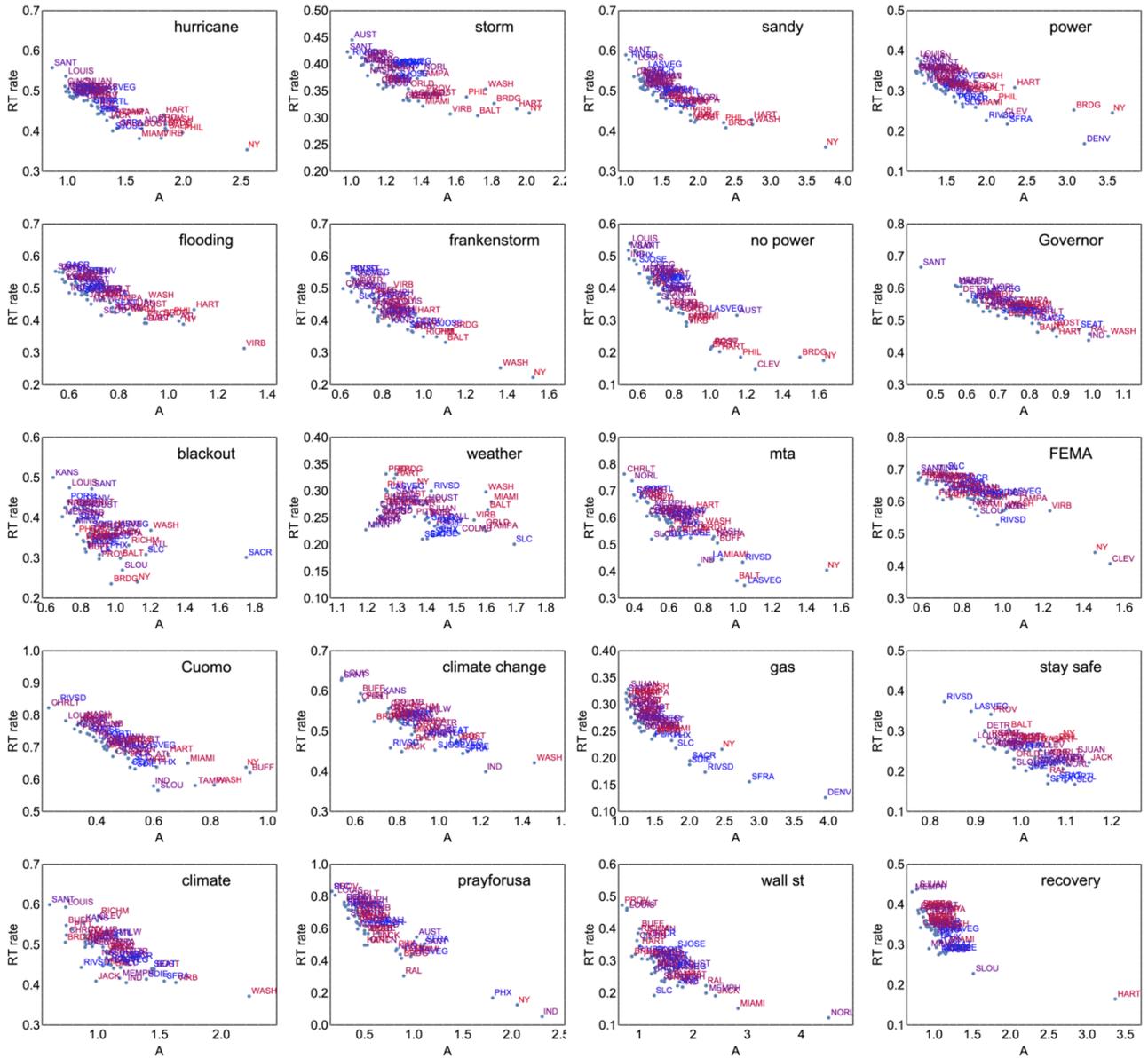

**Figure S2** Originality of the content, expressed through the fraction of re-tweets in the stream of messages. Most of the words show the inverse relationship between normalized activity and re-tweet rate; however, in the event-related keywords this trend is more pronounced (compare "sandy" or "frankenstorm" with "weather" or "climate").

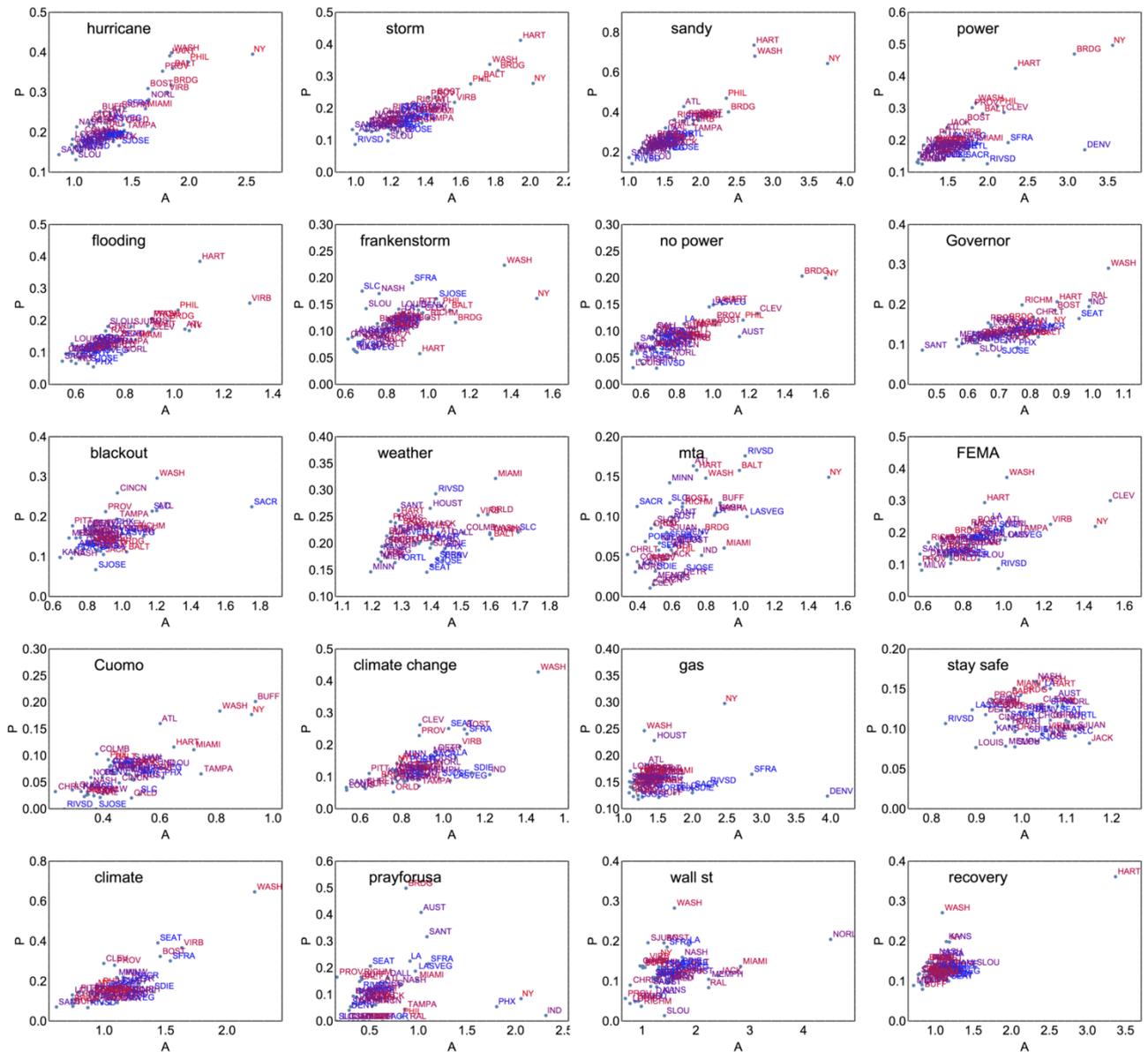

**Figure S3** Global popularity of local content. Words are sorted, as in Figure S1, according to relevance based on the strength of activity-distance correlation for East Coast cities. Event-related keywords show a strong linear relationship between activity and global popularity of messages generated locally. As relevance of the word to the disaster weakens, so does the correlation between activity and popularity.

**Table S2** Ranking of the keywords included into analysis according to strength of the correlation between the distance and activity for East Coast cities. Event- and effect-related words on the top are used for further analysis of activity, including its relationship to damage.

|  | Kendall rank | | Spearman rank | |
|---|---|---|---|---|
| keyword | $\tau$ | P-value | $\rho$ | P-value |
| hurricane | -0.690 | $1.07\ 10^{-9}$ | -0.851 | $1.28\ 10^{-11}$ |
| storm | -0.644 | $1.23\ 10^{-8}$ | -0.8240 | $2.02\ 10^{-10}$ |
| sandy | -0.596 | $1.38\ 10^{-7}$ | -0.778 | $9.26\ 10^{-9}$ |
| power | -0.548 | $1.29\ 10^{-6}$ | -0.735 | $1.5\ 10^{-7}$ |
| flooding | -0.548 | $1.29\ 10^{-6}$ | -0.731 | $1.84\ 10^{-7}$ |
| frankenstorm | -0.533 | $2.42\ 10^{-6}$ | -0.71 | $5.82\ 10^{-7}$ |
| no power | -0.519 | $4.46\ 10^{-6}$ | -0.673 | $3.74\ 10^{-6}$ |
| Governor | -0.477 | $2.54\ 10^{-4}$ | -0.651 | $9.46\ 10^{-6}$ |
| blackout | -0.343 | 0.002 | -0.470 | 0.003 |
| weather | -0.272 | 0.016 | -0.382 | 0.018 |
| mta | -0.231 | 0.042 | -0.365 | 0.024 |
| FEMA | -0.215 | 0.058 | -0.306 | 0.062 |
| Cuomo | -0.179 | 0.113 | -0.284 | 0.084 |
| climate change | -0.084 | 0.458 | -0.089 | 0.596 |
| gas | -0.081 | 0.474 | -0.134 | 0.422 |
| stay safe | -0.081 | 0.474 | -0.122 | 0.467 |
| climate | -0.081 | 0.474 | -0.101 | 0.545 |
| prayforusa | -0.05 | 0.66 | -0.091 | 0.585 |
| wall st | 0.044 | 0.697 | 0.087 | 0.604 |
| recovery | -0.038 | 0.734 | -0.038 | 0.822 |

**Table S3** Activity-damage correlations across keywords, in order of decreasing strength. Note that event-related keywords (on the basis of activity-distance relationship) are most predictive of damage. For the final analysis we use the following selection: "sandy", "hurricane", "storm", "power", and "flooding" ("mta" is excluded because of insufficient number of active ZCTAs)

|  |  | Kendall rank | | Spearman rank | | Pearson | |
|---|---|---|---|---|---|---|---|
| keyword | active ZCTAs | $\tau$ | P-value | $\rho_S$ | P-value | $\rho_P$ | P-value |
| sandy | 517 | 0.33 | $4.3\ 10^{-29}$ | 0.47 | $2.2\ 10^{-29}$ | 0.5 | $2.4\ 10^{-34}$ |
| flooding | 169 | 0.27 | $1.3\ 10^{-7}$ | 0.4 | $7.\ 10^{-8}$ | 0.37 | $5.5\ 10^{-7}$ |
| power | 532 | 0.26 | $8.1\ 10^{-20}$ | 0.38 | $1.2\ 10^{-19}$ | 0.42 | $5.2\ 10^{-24}$ |
| hurricane | 494 | 0.26 | $6.2\ 10^{-18}$ | 0.37 | $1.3\ 10^{-17}$ | 0.38 | $1.\ 10^{-18}$ |
| mta | 65 | 0.24 | $5.3\ 10^{-3}$ | 0.34 | $5.1\ 10^{-3}$ | 0.36 | $3.7\ 10^{-3}$ |
| no power | 450 | 0.21 | $2.3\ 10^{-11}$ | 0.3 | $4.7\ 10^{-11}$ | 0.31 | $9.3\ 10^{-12}$ |
| storm | 478 | 0.21 | $1.4\ 10^{-11}$ | 0.3 | $1.6\ 10^{-11}$ | 0.31 | $3.5\ 10^{-12}$ |
| gas | 505 | 0.18 | $6.2\ 10^{-10}$ | 0.28 | $2.3\ 10^{-10}$ | 0.32 | $9.\ 10^{-14}$ |
| blackout | 207 | 0.18 | $7.9\ 10^{-5}$ | 0.27 | $1.\ 10^{-4}$ | 0.31 | $6.5\ 10^{-6}$ |
| climate | 97 | 0.19 | $6.\ 10^{-3}$ | 0.27 | $8.6\ 10^{-3}$ | 0.28 | $6.2\ 10^{-3}$ |
| Governor | 282 | 0.17 | $3.1\ 10^{-5}$ | 0.25 | $2.9\ 10^{-5}$ | 0.29 | $8.\ 10^{-7}$ |
| FEMA | 272 | 0.16 | $6.5\ 10^{-5}$ | 0.23 | $9.5\ 10^{-5}$ | 0.24 | $8.4\ 10^{-5}$ |
| prayforusa | 4 | 0. | 1. | 0.2 | $8.\ 10^{-1}$ | 0.52 | $4.8\ 10^{-1}$ |
| frankenstorm | 52 | 0.13 | $1.7\ 10^{-1}$ | 0.2 | $1.6\ 10^{-1}$ | 0.094 | $5.1\ 10^{-1}$ |
| climate | 128 | 0.13 | $2.6\ 10^{-2}$ | 0.19 | $2.9\ 10^{-2}$ | 0.2 | $2.\ 10^{-2}$ |
| recovery | 255 | 0.12 | $3.3\ 10^{-3}$ | 0.18 | $3.6\ 10^{-3}$ | 0.2 | $1.1\ 10^{-3}$ |
| stay safe | 209 | 0.097 | $3.7\ 10^{-2}$ | 0.15 | $3.6\ 10^{-2}$ | 0.16 | $1.7\ 10^{-2}$ |
| Cuomo | 71 | 0.071 | $3.8\ 10^{-1}$ | 0.13 | $2.9\ 10^{-1}$ | 0.32 | $5.7\ 10^{-3}$ |
| weather | 476 | 0.067 | $2.8\ 10^{-2}$ | 0.1 | $2.7\ 10^{-2}$ | 0.17 | $1.8\ 10^{-4}$ |
| wall st | 83 | 0.067 | $3.7\ 10^{-1}$ | 0.096 | $3.9\ 10^{-1}$ | 0.08 | $4.7\ 10^{-1}$ |

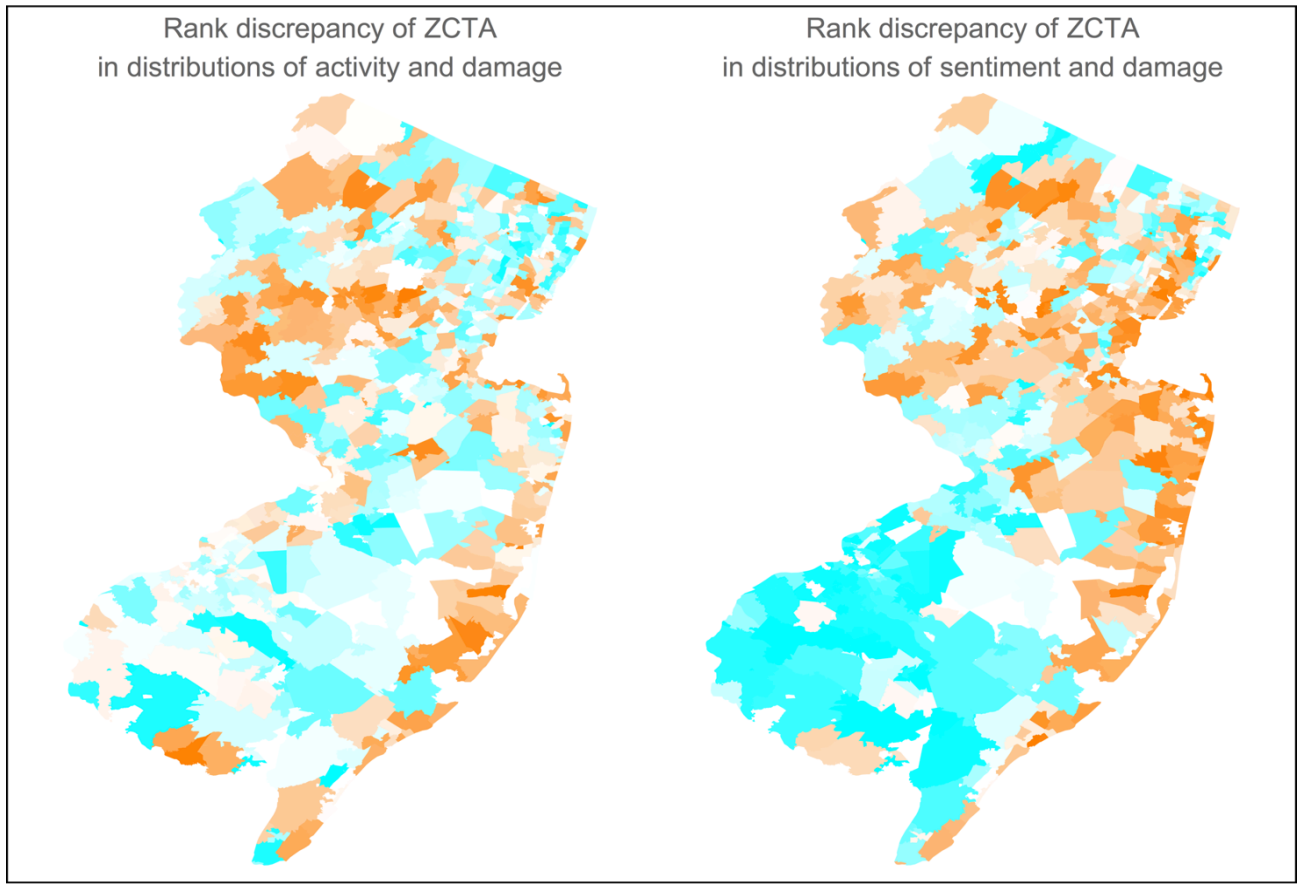

**Figure S4** Comparison of predictive capacity of activity and sentiment. The figure maps all zipcode tabulation areas shaded by color, with saturation reflecting discrepancy of the area's rank in two corresponding distributions. For instance, if a particular zipcode is 5th in the ranking of activity, but 100th in the ranking of damage the discrepancy is equal to 95. Discrepancies are normalized by the maximum observed deviation. The stronger the correlation is between the distributions the more uniform and light the map would be, as is the case for activity-vs.-damage map on the left.

**Table S4** Effect of normalization variable choice on the strength of activity-damage relationship (ZCTA-resolution)

| | Variables normalized by … | | | |
| | … Census population | | … "Twitter population" | |
| **Correlation measure** | statistic | P-value | statistic | P-value |
| --- | --- | --- | --- | --- |
| Kendall $\tau_K$ | 0.32 | $2.58\ 10^{-29}$ | 0.32 | $8.35\ 10^{-29}$ |
| Spearman $\rho_S$ | 0.46 | $1.92\ 10^{-29}$ | 0.45 | $2.92\ 10^{-29}$ |
| Pearson $\rho_P$ | 0.49 | $8.18\ 10^{-34}$ | 0.49 | $6.7\ 10^{-34}$ |

**Table S5** County level estimates of damage: modeling (Hazus-MH) and *ex-post* data on insurance and FEMA Individual Assistance grants.

| County | Population | Tweets | Users | Damage estimates | |
| --- | --- | --- | --- | --- | --- |
| | | | | *ex-post*, $M | Hazus-MH, $M |
| Atlantic | 275422 | 1580 | 574 | 954 | 1630 |
| Bergen | 918888 | 9516 | 2727 | 729 | 1070 |
| Burlington | 451336 | 1684 | 670 | 54.6 | 164 |
| Camden | 513539 | 1004 | 588 | 147 | 103 |
| Cape May | 96304 | 997 | 331 | 29.3 | 740 |
| Cumberland | 157785 | 521 | 265 | 12.7 | 128 |
| Essex | 787744 | 8260 | 1908 | 844 | 375 |
| Gloucester | 289586 | 1106 | 470 | 6.29 | 151 |
| Hudson | 652302 | 9322 | 2140 | 314 | 3600 |
| Middlesex | 823041 | 8070 | 2102 | 406 | 776 |
| Monmouth | 629384 | 8246 | 1865 | 919 | 1930 |
| Ocean | 580470 | 4404 | 1052 | 587 | 3240 |
| Passaic | 502885 | 3840 | 1237 | 41.8 | 34.2 |
| Salem | 65774 | 122 | 92 | 18.6 | 167 |
| Union | 543976 | 5946 | 1360 | 87.2 | 395 |
| Bronx | 1408473 | 2459 | 944 | 50.6 | 635 |
| Kings | 2565635 | 10040 | 3111 | 660 | 5470 |
| Nassau | 1349233 | 9085 | 2363 | 1590 | 6860 |
| New York | 1619090 | 50767 | 15558 | 252 | 4820 |
| Orange | 374512 | 1310 | 608 | 39.2 | 22.7 |
| Putnam | 99607 | 568 | 218 | 0.2 | 0.405 |
| Queens | 2272771 | 9453 | 2662 | 832 | 3650 |
| Richmond | 470728 | 3538 | 699 | 353 | 1880 |
| Rockland | 317757 | 2046 | 509 | 83.3 | 86.8 |
| Suffolk | 1499273 | 11851 | 3119 | 569 | 2720 |
| Ulster | 181791 | 400 | 233 | 0.524 | 8.03 |
| Westchester | 961670 | 6347 | 2234 | 237 | 1320 |

**Table S6** Strength of activity-damage correlations for different damage estimates

| | Damage is estimated … | | | |
| --- | --- | --- | --- | --- |
| | … by modeling (Hazus-MH) | | … from insurance and FEMA claims | |
| Correlation measure | statistic | P-value | statistic | P-value |
| Kendall $\tau_K$ | 0.29 | 0.035 | 0.34 | 0.013 |
| Spearman $\rho_S$ | 0.45 | 0.020 | 0.50 | 0.007 |
| Pearson $\rho_P$ | 0.37 | 0.056 | 0.40 | 0.036 |

**Table S7** Predictive power of sentiment, analyzed at different spatial resolutions and normalized either by the area Census population or local Twitter user count ("Twitter population")

| | Kendall τ for sentiment-damage relationship | | | |
|---|---|---|---|---|
| | ZCTA | | County | |
| **Normalization** | **statistic** | **P-value** | **statistic** | **P-value** |
| Census population | -0.029 | 0.31 | -0.19 | 0.11 |
| Twitter users count | -0.07 | 0.014 | -0.23 | 0.06 |